\documentclass[pre,manuscript]{revtex4}

\usepackage{graphicx} \usepackage{times} \usepackage{color}

\usepackage{amssymb,amsfonts,amsmath}

\begin{document} \title{\textbf{\huge The Effect of Recombination on
the Neutral Evolution of Genetic Robustness }} \author{Gergely
J. Sz\"oll\H{o}si} \email[]{ssolo@angel.elte.hu}
\homepage[]{angel.elte.hu/~ssolo} \author{Imre Der\'enyi}
\email[]{derenyi@angel.elte.hu} \homepage[]{angel.elte.hu/~derenyi}
\affiliation{Biological Physics Department, E\"otv\"os University,
Budapest}

\begin{abstract}

 Conventional population genetics considers the evolution of a limited
number of genotypes corresponding to phenotypes with different
fitness.  As model phenotypes, in particular RNA secondary structure,
have become computationally tractable, however, it has become apparent
that the context dependent effect of mutations and the many-to-one
nature inherent in these genotype-phenotype maps can have fundamental
evolutionary consequences. It has previously been demonstrated that
populations of genotypes evolving on the neutral networks
corresponding to all genotypes with the same secondary structure only
through neutral mutations can evolve mutational robustness [Nimwegen
{\it et al.} Neutral evolution of mutational robustness, 1999 PNAS],
by concentrating the population  on regions of high neutrality.
Introducing recombination we demonstrate, through numerically
calculating the stationary distribution of an infinite population on
ensembles of random neutral networks that mutational robustness is
significantly enhanced and further that the magnitude of this
enhancement is sensitive to details of the neutral network topology.
Through the simulation of finite populations of genotypes evolving on
random neutral networks and a scaled down microRNA neutral network, we
show that even in finite populations  recombination will still act to
focus the population on regions of locally high neutrality.

\end{abstract}
\maketitle

\section{introduction}

Robustness is the invariance of phenotypes in the face of perturbation
\cite{perspective}.  It has been found to be present in living
organisms at all levels of organizational complexity ranging from
protein structure to small genetic circuits \cite{wagner_plos,vulva},
genome-scale networks such as metabolic networks
\cite{kacser,barabasi} and even entire organisms.  The evolutionary
mechanism for the observed robustness has, however, been far from
clear. The fundamental problem lies in determining whether robustness
is an evolved property or intrinsic to genetic systems
\cite{perspective}.  An important stepping stone  in solving this
problem is establishing  a theoretical understanding of under what
conditions and to what extent natural selection can lead to the
evolution of robustness.

According to the classic results of Kimura and Maruyama the average fitness of an asexually reproducing population (in the limit of very large populations) depends only on the mutation rate and is independent of the details of the fitness landscape \cite{Kimura66}. This result, however, as stated in their paper, only holds under the assumption that the fittest genotype does not have any neutral sites. If neutral mutations are taken into account the average fitness of the population will depend on both the mutation rate and the details of the fitness landscape \cite{wilke}. A growing body of work has explored the 
effect of neutral mutations on mutation-selection balance in infinite 
populations (quasispecies) and the balance of mutation-selection and
drift in finite populations. Quaispecies theory \cite{wilke} and
simulations of finite populations of genotypes evolving with phenotype
defined by RNA secondary structure \cite{wagner_jzoo,huynen} or simple
lattice models of protein folding \cite{bornberg} have established
that a selective pressure to evolve robustness against mutations
exists. The net effect of this selection pressure is to concentrate the population in regions of genotype space where the density of neutral sequences is higher, selecting individual sequences with an increased robustness against mutations. Focusing on the case where all mutations are either neutral or
lethal, van Nimwegen {\it et al.} \cite{nimwegen} were able to solve
the quasispecies equations and have demonstrated that the extent to
which mutational robustness (the average number of neutral neighbors)
evolves depends solely on the topology of the network of neutral
genotypes.

In this work we examine the effects of recombination on the evolution
of mutational robustness on networks of neutral genotypes and its
dependence on the topology of the network, both in the limit of
infinite populations and for finite populations. Previous work on the
effects of recombination on the population dynamics on networks of
neutral genotypes is scarce, Xia {\it et al.} \cite{xia} have shown
that in a simple lattice model of protein folding recombination leads
to increased thermodynamic stability, but have not  addressed
the evolution of mutational robustness. In a wider context the effects
of recombination on the evolution of robustness has been studied in
highly simplified model developmental networks \cite{azevedo}, where
simulation results suggest that recombination between model gene
networks imposes selection for mutational robustness, and that
negative epistasis evolves as a byproduct of this selection.  Even
though   the evolution of mutational robustness in network of neutral
genotypes under recombination has not been previously studied in
detail the observation that recombination has a contracting property
(i.e.\ it always creates genotypes that are within the boundaries of
the current mutational cloud) has lead to the expectation that it
should concentrate the population in those regions of genotype space
in which the density of neutral genotypes is highest \cite{wilke}.

In order to gauge the most general effects of recombination we define
ensembles of random networks of neutral genotypes and demonstrate that
recombination leads to the concentration of the population in highly
neutral region of the genotype space, and hence the evolution of
mutational robustness, in these generic neutral networks.  Turning to
finite populations we show that similar to the case where only
mutation is present (cf. \cite{nimwegen}) the  evolution of mutational
robustness requires the population to be sufficiently polymorphic
(i.e.\ the product of the population size and the mutation rate, $\mu
N$, must be greater then one). We also demonstrate using a scaled down
analog of microRNA stem-loop hairpin structures that, provided a
sufficiently polymorphic population, recombination will lead to the
evolution of mutational robustness on larger and more realistic
neutral networks by concentrating the population on local regions of
genotype space with high neutrality.

\section{Model}
 
\subsection{Population dynamics}

Considering the genotype space of sequences of length $L$ over some
alphabet $\mathcal A$, following van Nimwegen {\it et al.}
\cite{nimwegen} we assume that genotype space contains a neutral
network of high, but equal fitness genotypes -- those sharing a common
preferred phenotype.  We assume, further, that the majority of the
population is concentrated on this neutral network, i.e.\ that the
remaining part of genotype space consist of genotypes with markedly
lower fitness, such that to good approximation all such phenotypes
maybe considered lethal.  For our evolutionary process we assume a
selection-mutation-recombination dynamics with constant population
size $N$. We consider all mutations or recombination events leading
off the neutral subset of genotypes $G$ with high fitness to be
fatal. Each individual of the population suffers mutations at a rate
$\mu L$ and undergoes recombination with a random member of the
population at a rate $\rho L$. Individuals who acquire a fatal
genotype (one not part of the the neutral network) are replaced
through reproduction of a random individual of the population.

 In the limit of large populations the stationary distribution of the
population on a given neutral network $G$  only depends on the ratio
$r=\rho/\mu$ and the neutral network $G$.   Considering one-point
recombination crossover events and denoting the number of neutral
genotypes in $G$ by $M$ and the frequency of genotype $i \in \{0,M\}$
in the population by $x_i$ the time evolution of our system in the
$N\to \infty$ limit is given by:
\begin{equation} \dot x_i = \sigma x_i- (\mu L + \rho L ) x_i + \mu
\sum_{j=1}^{M} M_{ij} x_j +  \rho \sum_{k=1}^{M}\sum_{l=1}^{M}
R_{ikl}  x_k x_l,   
\label{det_dyn}
\end{equation}     where $M_{ij} = 1$ if genotype $i$ can be derived
from genotype $j$ by replacing a single letter and zero otherwise,
$R_{ikl}$ equals the number of recombination (one-point crossover)
events through which genotypes $k$ and $l$ yield genotype $i$, and 
\begin{equation} \sigma= (\mu L + \rho L )  - \sum_{i=1}^{M} \left (
\mu \sum_{j=1}^{M} M_{ij} x_j + \rho \sum_{k=1}^{M}\sum_{l=1}^{M}
R_{ikl} x_k x_l \right ). 
\end{equation}  is the uniform growth rate of the
population, that compensates for the disappearance of lethal genotypes generated by
mutation and recombination, ensuring that the population size remains
constant, i.e. $\sum_i \dot x_i=0$ at all times. We compute the
limit distribution of the population ($x^{\rm st.}_i$) by numerically
computing the stationary solution of equations (\ref{det_dyn}).    

For finite $N$ stochasticity resulting from the discrete nature of the
reproduction process must also be taken into consideration.
Considering discrete generations we may proceed  by solving equation
(\ref{det_dyn}) while introducing sampling noise (i.e.\ drift) by
sampling the population at unit time intervals, the mutation and
recombinations rates will then be in units of events per generation.
Sampling consists of choosing $N$ individuals and assigning each a
genotype $i$ with probability $x_i$ and subsequently updating $x_i$
accordingly and integrating equation (\ref{det_dyn}) for one
generation time before performing sampling again. The resulting
stochastic population dynamics depends on the product of the mutation
rate and population size $\mu N$, the ratio of the mutation and
recombination rates $r$, and the structure of $G$.         
 For sufficiently large sample size the stochastic dynamics of our model only depends on the combined parameters $\mu N$ and $r$. Appropriately large sample size $N$ was chosen by increasing $N$   
(while keeping $\mu N$ and $r$ fixed) until the effects of sample size became negligible ($N \lessapprox 1000$).

Computing $R_{ikl}$ requires enumerating all $M \times M \times L$
possible recombination events.  This is only feasible for $M$ not
larger then a few thousand genotypes. For larger neutral networks
calculation of the infinite population limit stationary distribution
is not currently tractable, it is still possible, however, to simulate
the finite population dynamics as the complexity of these computations
scales with $\mathcal O (N^2 L)$ and not $\mathcal O (M^2 L)$. In
these simulations we average over several runs starting form random
initial conditions.

\subsection{Random neutral networks}

We investigated the effects of recombination on two types of random
neutral network ensembles. In both cases we choose an alphabet of size
$4$ and generated random neutral networks consisting of $M$ genotypes
of length $L$. The first type of random networks (hereafter referred
to as uniform attachment networks) were generated by choosing a random
seed genotype and adding a random neighboring genotype (one that can
be obtained from it by a single mutation). This process was continued
by selecting a random genotype from those already added to the network
and adding one its neighbors not already in the network at random
until the network contains $M$ genotypes. The second type of random
networks (hereafter referred to as preferential attachment networks)
were grown at each step by choosing genotypes from those already added
to the network with a probability proportional to the number of
neighbors the genotype already has in the network, and adding a
neighbor not contained in the network at random until the network
contained $M$ genotypes.

This choice of random ensembles is motivated by the expected
differences in the topology of networks  belonging to the two
ensembles. Preferential networks have a topology where the most
connected genotypes are also more centrally located on average.

\subsection{Scaled down microRNA neutral network}

The structure of microRNA precursor stem-loops, have been shown to
exhibit a significantly high level of robustness in comparison with
random RNA sequences with similar stem-loop structures
\cite{bornestein}.  This observation makes the neutral networks 
with identical stem-loop like secondary structure a natural testing
ground for the evolution of mutational robustness. To construct a
neutral network on which the above population dynamics is
computationally tractable  we proceed to construct a scaled down
analog of such stem-loop structures. We downloaded all currently
available miRNA stem-loop precursor sequences from miRBase
\cite{miRBase}. Analysis of the sequences and the corresponding
structures indicated that hairpin like stem-loops consisted of, on
average, a $7.241$-base-long loop and a $50.429$-base-pair-long stem
region.  To first approximation we may consider the neutral network of
a hairpin like structure to consist of large quasi-independent regions
corresponding to mutations in the central loop region that are
connected by more rare mutations in the stem region. Aiming to
approximate the neutral networks of stem-loop structures by that of a
single such region we  used the Vienna RNA secondary structure
prediction package \cite{vienna} to find connected neutral networks (a
set of genotypes that can be reached through single mutations) of the
hairpin like structure with $3$-base-pair-long stem and a $7$-base-long
loop region. In the following we present results for the $M=37972$
connected neutral network that contains the sequence GACUCGCACUGUC.     

\section{Results}
\subsection{ The effects of recombination in the infinite population
limit}

van Nimwegen {\it et al.} \cite{nimwegen} have shown that in the limit
of large populations the average number of neutral single mutant
neighbors in a population in selection-mutation balance  is larger
then the average number of neutral neighbors in the network. The
population tends to concentrate in  parts of the network with enhanced
neutrality. Introducing recombination and     numerically computing
the stationary distribution of the population in
selection-mutation-recombination balance using equation
(\ref{det_dyn}) we observe that recombination concentrates  the
population even further. This observation can be quantified by
comparing the entropy of the stationary population  distributions
$x^{\rm st.}_i$ defined by:
\begin{equation} H = \ln \frac{1}{M}   - \sum_{i=1}^M x^{\rm st.}_i
\ln x^{\rm st.}_i,
\end{equation}    for different values of $r$.  To asses the
mutational robustness of the population we compute the average number
of neutral neighbors of a random individual in the population:
\begin{equation} D = \sum_{i=1}^M x^{\rm st.}_i d_i, 
\end{equation}    where $d_i$ is the number genotypes in the neutral
network that can be obtained from genotype $i$ by replacing a single
letter.  To obtain a measure of the extent to which excess mutational
robustness emerges solely as an effect of the population dynamics we
compare this value to the average neutrality of the network:
\begin{equation} D_0 = \frac{1}{M}\sum_{i=1}^M d_i. 
\end{equation}   

To compute the above averages we generated $10^5$ random neutral
networks with $M=200$ and $L=20$ of both types and averaged $H$ and
$D/D_0$ over them. We also generated networks with larger $M$ (and
$L$) values and found qualitatively similar results. We note that due
to the finite nature of both the mutation rate and population size in
natural populations (together quantified by $\mu N$) it is  networks
of relatively small $M$ that are most relevant biologically, in the
sense that natural populations with finite $\mu N$ are in effect
restricted to some relatively small region of the neutral network for
time scales long enough for local selection-mutation-recombination
balance to be achieved. This implies that the extent to which
robustness is evolved is determined by   the local topology. The
exploration of the entire neutral network occurs on a much longer time
scale. This long time scale exploration is of secondary importance
from the perspective of the evolution of mutational robustness as the
topology of the neutral network beyond the mutational cloud is in
effect invisible to the population dynamics.

As shown in Fig. 1.a. and Fig. 1.c. recombination leads to a similar
increase in entropy in both uniform attachment and preferential
attachment random networks.  The entropy increase as a function of $r$
is slightly less significant for  preferential attachment networks $ H
=0.726 \pm 0.128,\ 0.982\pm 0.14,\ 2.102 \pm 0.19,\ 3.379 \pm 0.143$
for $r=0,1,10,100$, respectively, then for uniform attachment
networks,  $ H =0.492 \pm 0.76,\ 1.449 \pm 0.202,\ 2.518\pm 0.207,\
3.541 \pm 0.219$ for $r=0,1,10,100$, respectively.  

If we look, however, at the increase in average mutational robustness
(Fig. 1.b. and Fig.1.d.) we find that  as a function of $r$ it is
significantly larger for preferential attachment networks $ D/D_0
=1.775 \pm 0.067,\ 2.200\pm 0.108,\ 3.079 \pm 0.268,\ 3.951 \pm 0.494$
for $r=0,1,10,100$, respectively, then for uniform attachment
networks,  $ D/D_0 =1.684 \pm 0.087,\ 2.024 \pm 0.125,\ 2.569\pm
0.260,\ 3.045 \pm 0.422$ for $r=0,1,10,100$, respectively. This
suggest that in preferential attachment networks, were the expected
neutrality of genotypes with high centrality reaches higher levels,
evolve higher levels of mutational robustness.

\subsection{ The effects of recombination in finite populations   } 

Similar to the case where only mutation is present the evolution of
mutational robustness in the presence of recombination requires that
the population be sufficiently polymorphic, i.e.\ that the product of
the mutation rate and the population be greater then unity. If,
however, this condition is satisfied the presence of recombination
leads to the evolution of increased mutational robustness under rather
general circumstances. To quantify the extent to which mutational
robustness increases we proceed by considering the time average over
the stochastic population dynamics.  Due to the separation of the time
scales at which the population attains local
selection-mutation-recombination balance and the much longer
time-scale over which it explores the entire neutral network,
averaging over a set of random initial conditions was used to attain a
computationally tractable approximation of the long time average. 

Performing simulations for different values of $\mu N $ shows that
$D/D_0$ approaches its  infinite population as $\mu N$ is increased
(see Fig.2.). This indicates that recombination concentrates the
population  on local regions of higher neutrality for smaller $\mu N$
as well.  In order to examine the effects of recombination on a more
realistic neutral network we also performed simulations on a scaled
down version of a microRNA stem-loop (Fig.3.). We found that for all
values of $\mu N$ the extent to which the population evolves
mutational robustness is higher in the presence of recombination then
in its absence.

\section{Conclusions} We examined the effects of recombination on the
extent to which populations evolving on neutral networks exhibit
mutational robustness. Our results show that recombination leads to
enhanced mutational robustness under very general
circumstances. Calculating the stationary limit distribution of
populations evolving on neutral networks drawn from two different
random ensembles in the infinite population limit indicate that
populations in which recombination is present are more
sensitive to details of the topology than populations where only
mutation is present. In particular,  neutral networks where genotypes
of high centrality exhibit larger neutrality evolve greater mutational
robustness.  

While our results indicate that significant mutational robustness
readily evolves in the presence recombination, this result must be
considered with the caveat that evolution of mutational robustness
through neutral dynamics requires sufficient polymorphism to be
present in the population. Our results shown, however,
that provided this condition is met,  recombination leads to
increased values of mutational robustness in comparison to populations
where only mutation is present.  

\section*{Acknowledgments}
This work was supported by the Hungarian Science Foundation (K60665).

\begin{figure}
\begin{center}
\centerline{\includegraphics[width=0.95\textwidth]{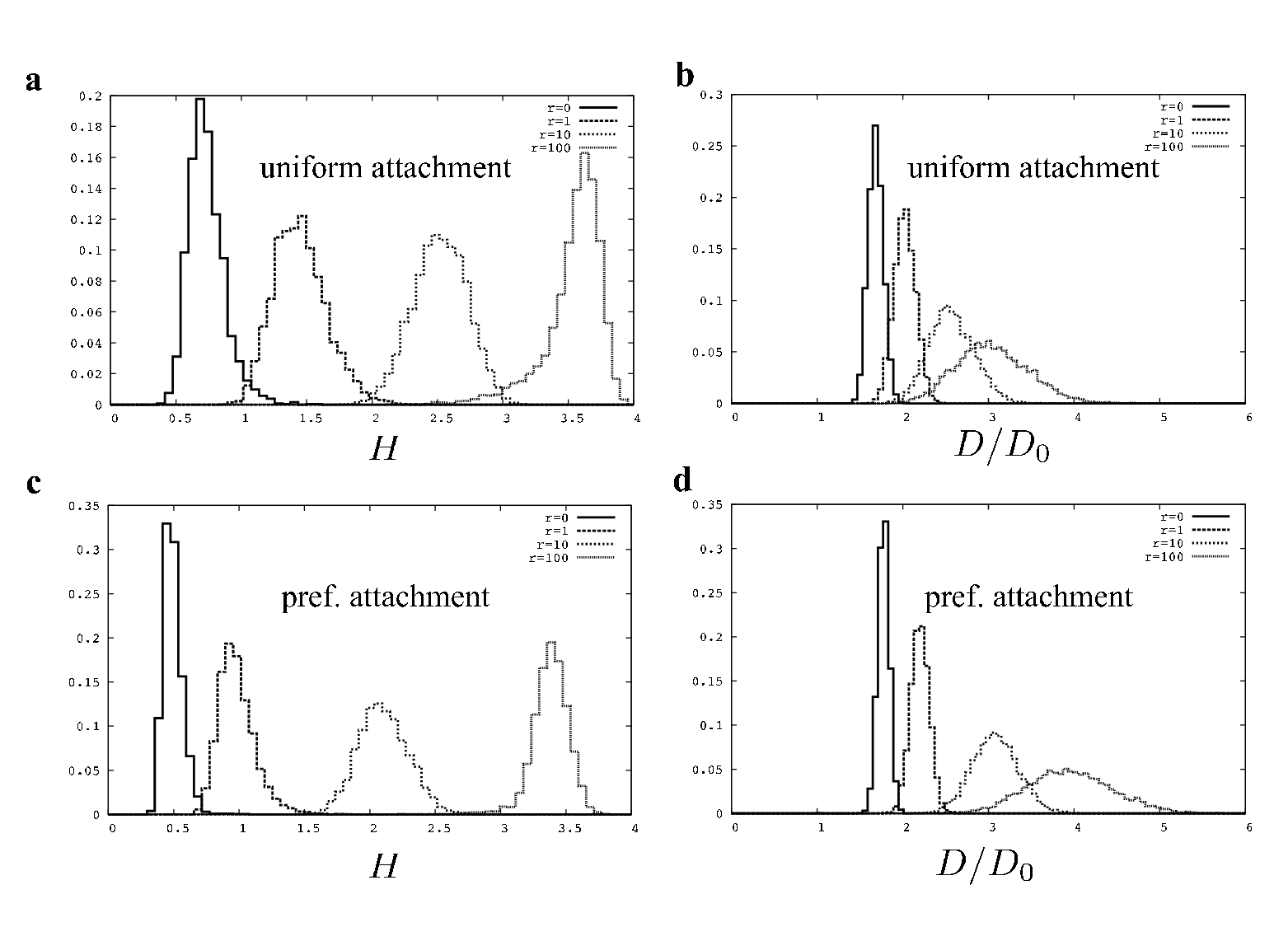}}
\caption{Histograms of the entropy $H$ and mutational robustness
enhancement  $D/D_0$ for different values of $r$. Numerically
calculating the stationary distribution of the population  on $10^5$
neutral networks $M=200$ genotypes of length $L=20$ randomly drawn
from the uniform attachment ensemble ({\bf a, b}) and preferential
attachment ensembles ({\bf b, c}) indicated that recombination leads
to significant enhancement of mutational robustness under very general
conditions.  Comparison of the results for the two ensembles suggests
that preferential attachment networks, where genotypes of higher
centrality are more neutral, evolve higher levels of mutational
robustness.}
\end{center}
\end{figure}

\begin{figure}
\begin{center}
\centerline{\includegraphics[width=0.95\textwidth]{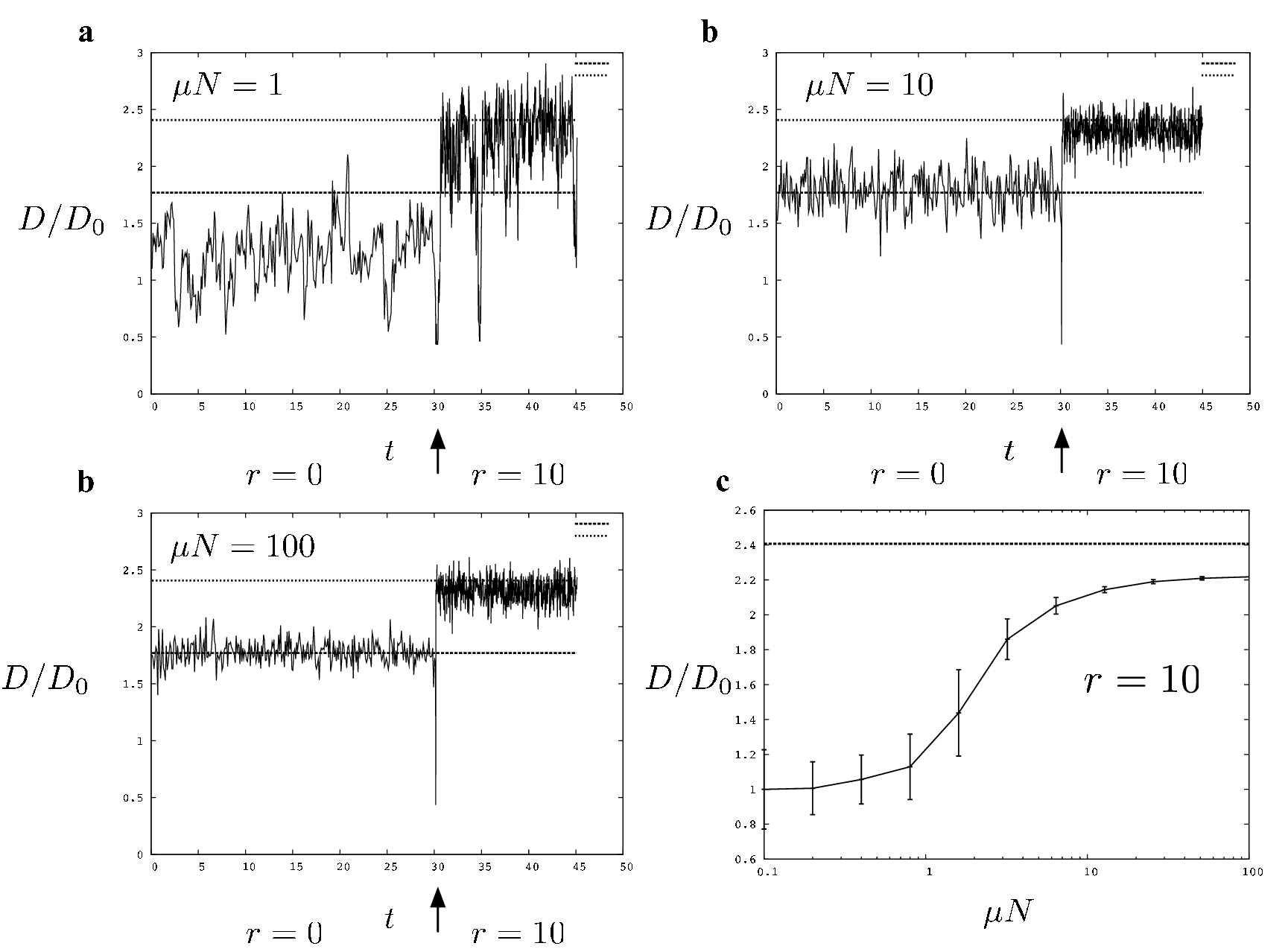}}
\caption{Simulations on a random uniform attachment network with
$M=200$, $L=20$ with different values of $\mu N$ show that the extent
to which mutational robustness is evolved increases as  $\mu N$
becomes larger. ({\bf a-c}) snapshots of the time evolution  of
mutational robustness enhancement $D/D_0$ in a population evolving  on
the same random uniform attachment network with different values of
$\mu N$, time is indicated in units of $2N$ generations. Recombination
was turned on at $t=30$ (indicated by the arrow). ({\bf d}) mutational
robustness enhancement $D/D_0$ as a function of $\mu N$ for the same
network  with $r=10$ as calculated from $100$ simulations with random
initial  conditions, where the population was allowed to evolve for
$100 \times 2 N$  generations, the error bars indicate the variance in
the time averages  over runs with random initial conditions. The
dashed line indicates the  value of the mutational robustness
enhancement $D/D_0$ in the $N \to \infty$  limit, throughout.}
\end{center}
\end{figure}

\begin{figure}
\begin{center}
\centerline{\includegraphics[width=0.75\textwidth]{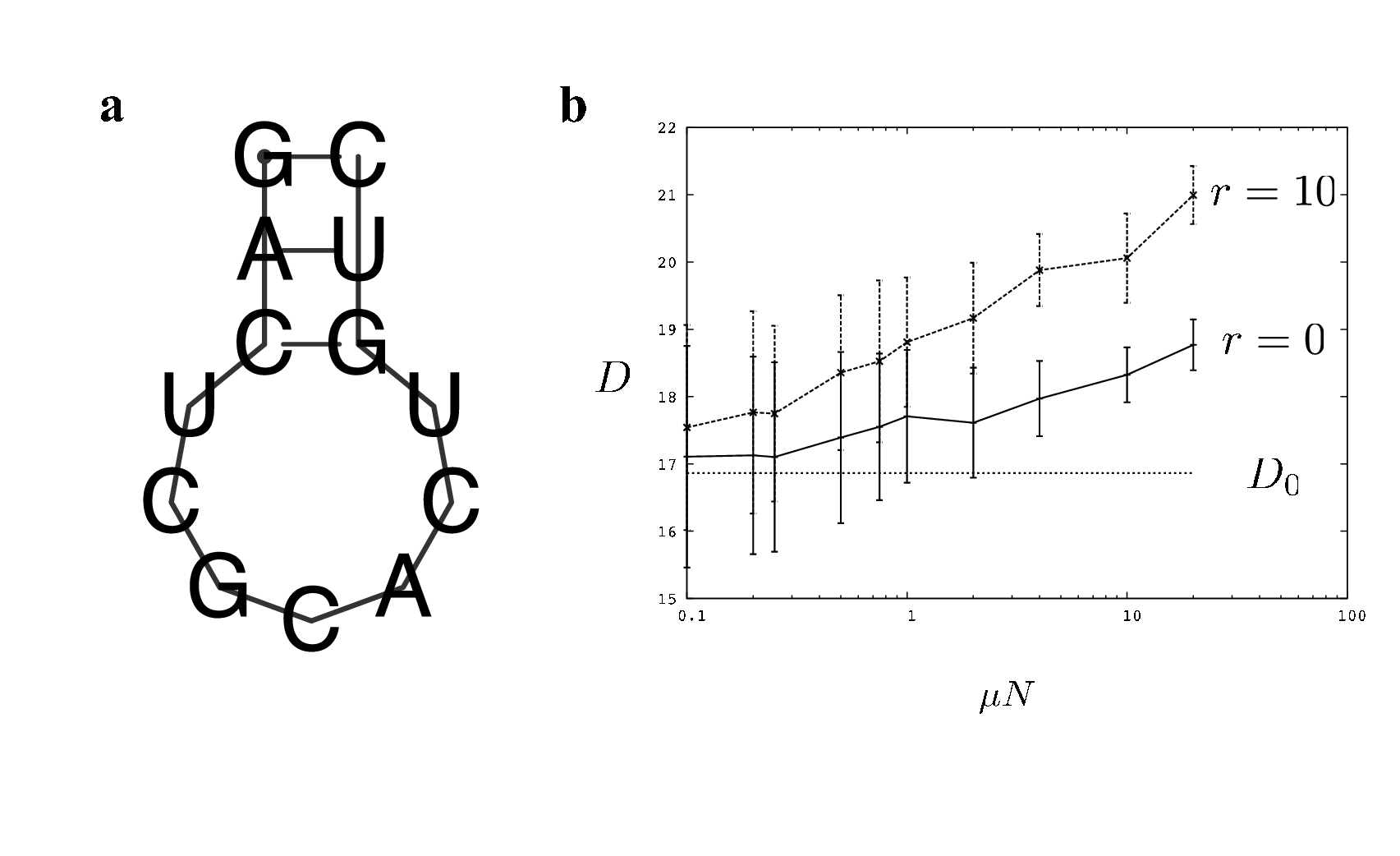}}
\caption{To investigate a more realistic neutral network we performed
simulations using a scaled down analog of microRNA stem-loop hairpin
structures ({\bf a}) consisting of a $7$ nucleotide long loop and a
$3$  nucleotide long stem region. ({\bf b}) The extent of mutational
robustness  was found to be higher in the presence of recombination
($r=10$) then without it ($r=0$) for all values of $\mu
N$. Simulations for different values  of $\mu N$ were performed for a
set of $20$ random initial conditions  starting from which the
population was allowed to evolve for  $100 \times 2 N$ generations,
the error bars indicate the variance in the  time averages over runs
with random initial conditions.  The dashed line  indicates the value
of $D_0$ throughout.}
\end{center}
\end{figure} 

\end{document}